\documentclass[11pt]{article}

\usepackage{SIGACTNews}

\newcommand{\COLUMNsplit}{\vskip 2em\hrule \vskip 3em}

\newcommand{\COLUMNguest}[2]{%
  \begin{center}%
   {\LARGE\bf #1 \par}%
    \vskip 1.5em%
    {\Large
      \lineskip .75em%
      \begin{tabular}[t]{c}%
        #2
      \end{tabular}\par}%
    \vskip 1.5em%
  \end{center}%
  \par}

\usepackage{graphicx}
\usepackage{fullpage}
\usepackage{amsmath}
\usepackage{amsfonts}
\usepackage{amssymb}
\usepackage{url}
\usepackage{float}%
\newtheorem{theorem}{Theorem}

\newtheorem{definition}{Definition}

\floatstyle{boxed}
\restylefloat{figure}

\title{SIGACT News Logic Column 13}
\author{Riccardo Pucella\\
Cornell University\\
Ithaca, NY 14853 USA\\
riccardo@cs.cornell.edu}
\date{}

\begin{document}

\SIGACTmaketitle

For this issue, Nick Papanikolaou brings us an introduction to
the subject of quantum logic and a survey of the relevant
literature, including a discussion of logics for
specification and analysis of quantum information
systems. 

I am always looking for contributions. If you have any suggestion
concerning the content of the Logic Column, or even better, if you
would like to contribute by writing a survey or tutorial on your own
work or topic related to your area of interest, feel free to get in
touch with me.

\COLUMNsplit

\COLUMNguest{Reasoning Formally about Quantum Systems: \\An Overview}
   {Nick Papanikolaou\\Department of Computer Science\\University of Warwick\\\texttt{http://www.warwick.ac.uk/go/nikos}}

\section{Introduction}

Quantum theory is widely accepted as the most successful theory of natural
science. Its precepts challenge our fundamental understanding of the universe,
and are often in direct conflict with what our intuition leads us to believe.
The implications of quantum theory for information processing are very hard to
ignore; indeed, to harness the potential of the quantum world is to enable
extremely powerful computational techniques, as well as novel means of data communication.

With the emergence of practical quantum cryptographic systems and related
products, there is already a growing need for means of designing and analysing
systems involving quantum--mechanical components. For instance, quantum
cryptographic protocols involve a sequence of steps for manipulating given
quantum states, and their implementation presupposes the existence of usable
quantum channels, with properties not found in conventional transmission
media; how is one to model protocols such as these and the properties they
exploit? How is one to demonstrate that a quantum--mechanical component
operates in accordance with its specification, or that it guarantees (where
applicable) a particular level of security?

Computer scientists have already developed a number of formalisms allowing one
to reason about quantum--mechanical behaviour in general as well as about
systems consisting of both conventional and quantum--mechanical elements. This
includes quantum programming languages (intended primarily for the description
of quantum algorithms; see the recent survey \cite{Gay} by S. Gay), quantum
process algebras (capable of describing quantum computational processes in
general, as well as quantum communication schemes), and logics for quantum
information systems.

The subject of `quantum logic' is an interesting line of work, reserved till
now for physicists and mathematicians who are interested in the algebraic
structures which arise in the mathematics of quantum theory. While it is a
very specialised subject, quantum logic can provide insights into the workings
of nature, and it has already been put to use by S. Abramsky and B. Coecke to
analyse problems in quantum information \cite{abramsky}.

This inquiry will be centred around the logical aspects of quantum theory. The
discussion will remain at a high level, focusing on philosophical problems and
the syntax of certain logics. First we will describe the principles of quantum
theory; we will pay attention particularly to the issue of quantum
measurement, which lies at the heart of Birkhoff's and von\ Neumann's `quantum
logic,' to which we will then turn our attention. This will be followed by a
summary of recent work by R. van der Meyden and M. Patra
\cite{meyden-know,Meyden2003,Patra,Patra2005}, and by P. Mateus and A.
Sernadas \cite{Mateus2004,Mateus2004a}, on logics for quantum information systems.

We have chosen to distinguish clearly between quantum logic and `logics for
quantum information systems;' the former is more of a semantic approach to
logic, with an emphasis on mathematical structures, while the latter refers to
logical calculi specifically targeted at applications in quantum computing and
quantum information theory. Quantum logic is usually studied at a very
abstract level, and applications or examples of it are quite superficial; our
presentation of quantum logic here is also abstract, and focuses on its
semantic connection with classical propositional logic. For the two other
logics, specific examples have been included.%

\paragraph{Acknowledgements.}
I wish to thank my supervisor, R. Nagarajan, for his guidance and particularly
for his patience with my independent writing efforts. I also thank
much--respected collaborator, S. Gay, and also my advisor, M. Jurdzi\'{n}ski,
for their counsel and encouragement.%

\paragraph{Disclaimer.}
I have received input on the material presented here from a number of people,
but any errors or omissions are entirely my own. I welcome any feedback; do
feel free to contact me at
\url{nikos@dcs.warwick.ac.uk}%
.

\section{Fundamentals of Quantum Theory}

A fundamental tenet of quantum theory is the belief that, at the lowest level,
the physical world is discrete, or \emph{quantised.} This claim is
corroborated by several experiments, described in all the standard texts on
the subject (see e.g. the books by Bohm \cite{Bohm1979}, Cohen--Tannoudji
\emph{et al.} \cite{Cohen-Tannoudji1977}, Peres \cite{Peres1995} and Shankar
\cite{Shankar1980}). More interestingly, quantum theory stipulates that there
is a limit to the amount of knowledge we can gain about a particular quantum
system; this is known as the \emph{uncertainty principle.} But the greatest
departure of quantum theory from the notions of classical physics is its use
of probability laws and, hence, the abandonment of causality. Quantum theory
thus presents us with several philosophical problems, and this directly
affects any attempt to arrive at its formulation in the language of logic. We
shall soon enter into a discussion of the details, but first a brief
introduction to the key aspects the theory is in order.

By the term \emph{physical system} is understood an identifiable, isolated
portion of the physical universe; such a system is characterised by its
\emph{state,} which is the result of experimental procedures used to isolate
and prepare it. An \emph{observable} is a quantity associated with the state
of a system, which can be directly measured. These are our basic terms of reference.

\subsection{The Hilbert Space Formalism\label{qu-formalism}}

Quantum theory finds mathematical expression in the so--called Hilbert space
formalism, which coordinates to a physical system a \emph{state space,}
namely, a complex--valued vector space $\mathcal{H}_{n}$\ equipped with an
inner product. The quantum state of a system is described by a vector in this
space; such a vector is normally written in the form $\left\vert
v\right\rangle $. Any physical system will have several degrees of freedom,
this being one of its intrinsic properties. The dimension, $n,$ of the state
space associated with the system reflects the number of degrees of freedom of
the system in question. Furthermore, every vector in this space is realisable
as an actual physical state.

The concept of quantum \emph{state} is very subtle and controversial; the
traditional view is that a quantum state represents all that can be known
about a system. This view is known to lead to some contradictions, so many
prefer to identify the concept of state not with an individual element of
reality, but with a description of ensembles of systems. If this view is
taken, then the most general form of quantum state is described by a
statistical operator, known as the \emph{density operator.} We will not adopt
this view here; rather, we will deal only with \emph{pure} quantum states, for
which the former view is satisfactory. The interested reader is referred to
advanced texts, such as \cite{Beltrametti1981,Ballentine1998}, for more details.

The state space $\mathcal{H}_{n}$\ of a given physical system may or may not
be finite--dimensional. It is only the former case with which we will be
concerned here.

Vectors in $\mathcal{H}_{n}$\ can be added together, producing so--called
\emph{superposition states}. It is a fundamental postulate of quantum
mechanics that, if $\left\vert v\right\rangle $ and $\left\vert w\right\rangle
$ represent valid physical states in $\mathcal{H}_{n}$, then so does their
superposition, i.e. the state $\alpha\cdot\left\vert v\right\rangle
+\beta\cdot\left\vert w\right\rangle $ with $\alpha,\beta$ being scalars. This
principle is put to great advantage in quantum computing, where it is used to
generate complex physical states with no classical analogue.%

\paragraph{Example.}
Consider the two--dimensional Hilbert space $\mathcal{H}_{2}$. A pair of
mutually orthogonal, normalised vectors, such as $\left\vert 0\right\rangle $
and $\left\vert 1\right\rangle ,$ forms a basis of $\mathcal{H}_{2}$. Another
basis of $\mathcal{H}_{2}$ is the pair of vectors
\[
\left\vert +\right\rangle =\frac{1}{\sqrt{2}}(\left\vert 0\right\rangle
+\left\vert 1\right\rangle ),\text{ \ }\left\vert -\right\rangle =\frac
{1}{\sqrt{2}}(\left\vert 0\right\rangle -\left\vert 1\right\rangle )
\]
\ The space $\mathcal{H}_{2}$ represents the state space, say, of a
spin-$\frac{1}{2}$ particle (where $\left\vert 0\right\rangle $ could denote
the particle's \textquotedblleft spin--down\textquotedblright\ state, and
$\left\vert 1\right\rangle $ its \textquotedblleft spin--up\textquotedblright%
\ state), or of a polarised photon (where $\left\vert 0\right\rangle $ would
stand for a polarisation angle of $0^{\circ}$ and 1 for a polarisation angle
of $90^{\circ}$). A more general, possible state of such a system is given by
a linear combination of the basis vectors, i.e. $\alpha\cdot\left\vert
0\right\rangle +\beta\cdot\left\vert 1\right\rangle $, for some $\alpha
,\beta\in%
\mathbb{C}
$. Quantum mechanics requires that $\left\vert \alpha\right\vert
^{2}+\left\vert \beta\right\vert ^{2}=1$. There is an infinity of such
combinations and thus, a quantum system with a state space even as simple as
this has infinitely more realisable states than its classical analogue. A
quantum system whose state space is specifically $\mathcal{H}_{2}$ is known as
a quantum bit or \emph{qubit.} Therefore, while a classical bit can only take
on a single value $b\in\{0,1\}$ at any given time, a qubit can be in the basis
states $\left\vert 0\right\rangle $, $\left\vert 1\right\rangle $, or any
superposition thereof.
\bigskip

The dual space $\mathcal{H}_{n}^{\mathcal{\ast}}$ of $\mathcal{H}_{n}$,
consists of linear functionals $\left\langle w\right\vert :\mathcal{H}%
_{n}\mapsto%
\mathbb{C}
$, where $%
\mathbb{C}
$ is the field of complex numbers. The \emph{inner product} $\left\langle
w|v\right\rangle $ induces an isomorphism between $\mathcal{H}_{n}%
^{\mathcal{\ast}}$ and $\mathcal{H}_{n}$; in particular, the image
$\left\langle v^{\prime}\right\vert \in\mathcal{H}_{n}^{\mathcal{\ast}}$\ of a
vector $\left\vert v\right\rangle \in\mathcal{H}_{n}$ is defined as a function
$f$ such that $f(\left\vert v\right\rangle )$ is the inner product of
$\left\vert v^{\prime}\right\rangle $ and $\left\vert v\right\rangle .$

\sloppypar A quantum system composed of multiple subsystems, whose state
spaces $\mathcal{H}_{k}^{(1)},\ldots,\mathcal{H}_{m}^{(N)}$ are known, has a
state belonging to the \emph{tensor product space,} written $\mathcal{H}%
^{\otimes n}=\mathcal{H}_{k}^{(1)}\otimes\cdots\otimes\mathcal{H}_{m}^{(N)}$.
The definition of the tensor product is detailed elsewhere, e.g.
\cite{Ballentine1998,Cohen-Tannoudji1977,nielsenchuang,rieffel}.

The state of a quantum system evolves over time, this evolution being governed
by a \emph{unitary transformation;} such a transformation is described by a
linear operator $U$ with $U^{-1}=U^{\dag}$. The symbol $U^{\dag}$ denotes the
\emph{adjoint} of operator $U$, defined as its conjugate-transpose: $U^{\dag
}=(U^{\ast})^{T}.$

In order to obtain any information about the quantum state of a system, a
measurement, or \emph{observation,} must be performed. An observable is
usually associated with an Hermitian operator, and the only possible results
of measurement are given by the eigenvalues of this operator. An operator $U$
is said to be \emph{Hermitian} in a finite--dimensional space if it is
self--adjoint, i.e. if $U=U^{\dag}.$ We will have more to say about the issue
of measurement in the following section.

\subsection{Quantum Measurement\label{sec:meas}}

Quantum theory makes of the measurement, or observation, of a quantum system
an issue of great importance; according to the theory, the actual state of a
particular quantum system cannot be determined experimentally, for there is an
interaction between the system and the means of observation. Indeed, this
interaction manifests itself as an irreversible disturbance to the state of
the system.

This result makes it difficult to reason about the actual quantum state of a
given physical system. One can only \emph{prepare} a system in a known state,
but any attempt to measure the state will directly affect it. Therefore, there
is a fundamental limit to the amount of information one can expect to obtain
about a given system. The best we can do is to predict, with particular
probability, the outcome of a specific measurement. Quantum mechanics provides
rules for calculating the probability distributions associated with the
measurement of system observables.

Interestingly, there are pairs of observables that are mutually dependent in
such a way that an accurate measurement of one precludes any reasonable amount
of accuracy in the measurement of the other, so that it is impossible to
properly measure both at the same time. The most typical example of this
arises when one tries to measure the position and momentum of a particle
simultaneously. It is not our object to expound further on this matter; the
reader should consult one of the several good texts on quantum theory.

For our purposes, it will suffice to say a few words about \emph{projective
measurements.} An observable described by an Hermitian operator, say $M$,
describes a projective measurement. When such a measurement is made on a
quantum system, the vector corresponding to its state is projected onto a
subspace of the state space. Furthermore, according to Nielsen and Chuang
\cite{nielsenchuang}:

\begin{quotation}
\noindent... \textsl{the possible outcomes of the measurement are the
eigenvalues of }$M$\textsl{. Upon making an observation with }$M$\textsl{\ of
the system in state }$\left\vert w\right\rangle $\textsl{, the probability of
getting an eigenvalue }$m$\textsl{\ is given by }$\left\langle w\right\vert
P_{m}\left\vert w\right\rangle $\textsl{\ where }$P_{m}$\textsl{\ is the
projector onto the eigenspace of }$M$\textsl{\ with eigenvalue }$m$\textsl{.
When the outcome }$m$\textsl{\ occurs, the quantum state evolves to the state
given by }%
\[
\frac{P_{m}\left\vert w\right\rangle }{\sqrt{\left\langle w\right\vert
P_{m}\left\vert w\right\rangle }}%
\]

\end{quotation}

Having established all the necessary background material, we are now ready to
start our journey into the logical aspects of quantum theory, and to enter
into a discussion of some of the logics which have been designed specifically
for reasoning about quantum systems.

\section{Quantum Logic}

The term \emph{quantum logic} is reserved for the study of the algebraic
structures which arise in the mathematical formalism of quantum mechanics. It
was G. Birkhoff and J. von Neumann who first pointed out that
quantum--mechanical \emph{propositions,} which are the simplest type of
observable associated with a given system, altogether constitute an
\emph{orthocomplemented, quasi--modular lattice}
\cite{Birkhoff1936,Beltrametti1981,Mittelstaedt1978}. This algebraic structure
has formal similarities with Boolean algebra, which provides the semantics of
classical propositional logic. In light of these similarities, Birkhoff and
von Neumann suggested that the lattice of all propositions associated with a
quantum system must provide the semantic foundation for a quantum--mechanical,
propositional calculus of logic; it is this particular calculus which was
christened `quantum logic.' In the words of Beltrametti \emph{et al.}
\cite{Beltrametti1981}:

\begin{quotation}
\textsl{Roughly, the starting question is whether the propositions of a
quantum system can be associated with, or can be interpreted as, sentences of
a language (or propositional calculus) and which rules this language inherits
from the ordered structure of propositions. In raising this question one has
in mind the fact that when the physical system is classical its propositions
form a Boolean algebra, and Boolean algebras are the algebraic models of the
calculus of classical logic. Thus, the question above can also be phrased as
follows: when a Boolean algebra is relaxed into an orthomodular
nondistributive lattice, which logic is it the model of? \textquotedblleft
Quantum logic\textquotedblright\ is the name that designates the answer, but
there are several views about the content of this name.}
\end{quotation}

The precise syntax and applicability of quantum logic has been debated for a
very long time and is still an active area of research. The fact that there is
currently no commonly agreed syntax for quantum propositions limits the
applicability of quantum logic to specific problems. But one should remember
that the emphasis in quantum logic is placed not on syntax, not on
applications, but on semantics. 

Let it be made clear at the outset that the subject of quantum logic is more
usually treated as part of a programme to understand quantum theory in depth.
We have the occasion to present this topic here, since it is likely to be of
some wider interest.

\subsection{Motivation for Quantum Logic}

As we have seen in Section \ref{sec:meas}, the formalism of quantum mechanics
provides us with a means of predicting the possible outcomes of a measurement
on a quantum system. These predictions have an associated probability of
materialising; if an actual measurement is made, only one of these predictions
will turn out to be correct. A quantum--mechanical\emph{ proposition }is just
such a prediction; quantum logic is the logic of these propositions. Birkhoff
and von Neumann \cite{Birkhoff1936}\ explain this as follows:

\begin{quotation}
\textsl{It is clear that an \textquotedblleft observation\textquotedblright%
\ of a physical system }$\mathcal{S}$\textsl{ can be described generally as a
writing down of the readings from various compatible measurements. Thus if the
measurements are denoted by the symbols }$\mu_{1},\ldots,\mu_{n},$\textsl{
then an observation of }$\mathcal{S}$\textsl{ amounts to specifying numbers
}$x_{1},\ldots,x_{n}$\textsl{ corresponding to the different }$\mu_{k}%
$\textsl{.}

\textsl{It follows that the most general form of a prediction concerning
}$\mathcal{S}$\textsl{ is that the point }$(x_{1},\ldots,x_{n})$\textsl{
determined by actually measuring }$\mu_{1},\ldots,\mu_{n},$\textsl{ will lie
in a subset }$S$\textsl{ of }$(x_{1},\ldots,x_{n})$\textsl{--space. Hence if
we call the }$(x_{1},\ldots,x_{n})$\textsl{--spaces associated with
}$\mathcal{S}$\textsl{, its \textquotedblleft
observation--spaces,\textquotedblright\ we may call the subsets of the
observation--spaces associated with any physical system }$\mathcal{S}%
$\textsl{, the \textquotedblleft experimental propositions\textquotedblright%
\ concerning }$\mathcal{S}$\textsl{.}
\end{quotation}

Quantum logic allows us to reason about measurements by defining logical
connectives for quantum--mechanical propositions. In terms of the
Hilbert--space formalism, a quantum--mechanical proposition is simply an
observable which admits two possible values (0 and 1).

The literature on the subject is generally concerned with the semantics of
quantum logic (see e.g.
\cite{Beltrametti1981,Birkhoff1936,chiaragiuntini,Mittelstaedt1978,Piron1976,Svozil1998,sep-qt-quantlog,coecke-book}%
). \ We will proceed to discuss the semantics of quantum logic after
explaining the connection between propositional logic and the theory of
lattices. Our treatment is basic, and we provide no proofs; mathematical
rigour is not our priority in this article.

\subsection{Boolean Algebra}

In formal logic, a proposition is an assertion which has a definite truth
value (either \textquotedblleft true\textquotedblright\ or \textquotedblleft
false\textquotedblright). The calculus of propositional logic allows to relate
propositions using connectives such as `and', `or' and 'not'. Each connective
expresses a \emph{logical operation;} for the connectives just mentioned, the
operations are, respectively, \emph{conjunction, disjunction} and
\emph{negation.} It was George Boole who realised that logical operations are
amenable to an algebraic treatment, and that they obey similar laws to
set--theoretic operations \cite{prest}. He extracted the algebraic laws for
these operations and came up with what is now known as `Boolean algebra.'
Boolean algebra is thus the mathematical structure common to the algebra of
sets and the algebra of propositions. The formal definition of a Boolean
algebra is given below.

\begin{definition}
[from \cite{prest}]A \textbf{Boolean algebra} $(B;\wedge,\vee,\lnot,0,1)$ is a
set $B$, together with operations on the set which satisfy certain laws. We
will denote the operations by `$\wedge$', `$\vee$', and `$\lnot$' and they
will be called `meet', `join' and `complement' respectively. There are two
distinguished and distinct elements of $B$, denoted by $0$ and $1$ that are
subject to the following laws.

\begin{enumerate}
\item Idempotence: $a\wedge a=a,$ and $a\vee a=a.$

\item Complement laws: $a\vee\lnot a=1,$ $a\wedge\lnot a=0,$ and $\lnot\lnot
a=a.$

\item Commutativity: $a\wedge b=b\wedge a,$ and $a\vee b=b\vee a.$

\item Associativity: $a\wedge(b\wedge c)=(a\wedge b)\wedge c,$ and
$a\vee(b\vee c)=(a\vee b)\vee c.$

\item Distributivity: $a\wedge(b\vee c)=(a\wedge b)\vee(a\wedge c),$ and
$a\vee(b\wedge c)=(a\vee b)\wedge(a\vee c).$

\item Property of 1: $a\wedge1=a.$

\item Property of 0: $a\vee0=a.$

\item De Morgan's laws: $\lnot(a\wedge b)=\lnot a\vee\lnot b,$ and
$\lnot(a\vee b)=\lnot a\wedge\lnot b.$
\end{enumerate}
\end{definition}

If we were to identify the Boolean operations, meet, join and complement with
the set--theoretic operations intersection `$\cap$', union `$\cap$' and
complement `$^{c}$', we would find that the above definition just gives the
laws of set theory, assuming $B$ is the set of all subsets of a given set $U$.
But the above definition also embodies the laws of propositional logic, which
is evident if we identify $B$ with the set of all equivalence classes of
propositions \cite{prest}.

A Boolean algebra is actually a special case of a more general algebraic
structure, known as a \emph{lattice.} In particular, a Boolean algebra is a
\emph{complemented distributive lattice,} defined as follows:

\begin{definition}
[from \cite{prest}]A \textbf{complemented distributive lattice} is a partially
ordered set $(P,\leqslant)$ that satisfies the following conditions:

\begin{enumerate}
\item Each pair of elements of $P$ has a least upper bound, denoted $a\vee b,$
and a greatest lower bound, denoted $a\wedge b$.

\item $(P,\leqslant)$ has a \emph{top} value, denoted $1$, and a \emph{bottom}
value, denoted $0$. The top value is defined as the element $c\in P$ which
satisfies $c\geqslant a$ for all $a\in P.$ Similarly, the bottom value is the
element $b\in P$ for which $b\leqslant a$ for all $a\in P.$

\item Distributivity holds in $P$, i.e. for any $a,b,c\in P,$ $a\wedge(b\vee
c)=(a\wedge b)\vee(a\wedge c),$ and $a\vee(b\wedge c)=(a\vee b)\wedge(a\vee
c).$

\item Every element in $P$ has a complement.
\end{enumerate}
\end{definition}

\begin{theorem}
[from \cite{prest}]Every complemented distributive lattice is a Boolean
algebra under the operations of meet, join and complement and with the
distinguished elements 0 and 1. Conversely, every Boolean algebra is a
complemented distributive lattice under the partial order that is defined by
\[
a\leqslant b\text{ if and only if }a=a\wedge b.
\]

\end{theorem}

The purpose of presenting these mathematical facts here is to demonstrate the
connection between logic and set theory. In order to understand Birkhoff and
von Neumann's work on quantum logic, one must be aware of the algebraic
structure of classical propositional logic.

\subsection{The Structure of Quantum--Mechanical Propositions}

Birkhoff and von Neumann describe their motivation and main result as follows
\cite{Birkhoff1936}:

\begin{quotation}
\textsl{The object of the present paper is to discover what logical structure
one may hope to find in physical theories which, like quantum mechanics, do
not conform to classical logic. Our main conclusion, based on admittedly
heuristic arguments, is that one can reasonably expect to find a calculus of
propositions which is formally indistinguishable from the calculus of linear
subspaces with respect to }\textit{set products, linear sums,}\textsl{ and
}\textit{orthogonal complements}\textsl{---and resembles the usual calculus of
propositions with respect to }\textit{and}\textsl{, }\textit{or}\textsl{ and
}\textit{not}\textsl{.}
\end{quotation}

As mentioned earlier, a quantum--mechanical proposition corresponds to an
observable with two possible values. As discussed in Section
\ref{qu-formalism}, an observable is described by a projection
operator\footnote{Not \emph{all} observables in quantum mechanics are
represented by projection operators, but we have restricted our discussion to
those which are (see Section \ref{sec:meas}).}, or \emph{projector.}
Therefore, every proposition about a given quantum system with state space
$\mathcal{H}$ has an associated projector. What is interesting is that the set
of all projectors $\mathcal{P(H)}$ on $\mathcal{H}$ is actually a lattice. We
will briefly list the characteristics of this structure here (see Beltrametti
\emph{et al. }\cite{Beltrametti1981} for details)\emph{.} First, the following
definitions are in order.

\begin{definition}
[from \cite{Piron1976}]A lattice $L$ is said to be \textbf{orthocomplemented}
if it is provided with an orthocomplementation, that is to say, a mapping of
$L$ onto $L$ which to each element $b\in L$ brings into correspondence an
element denoted as $b^{\prime}\in L$, such that:

\begin{enumerate}
\item $\forall b\in L:(b^{\prime})^{\prime}=b$

\item $\forall b\in L:b\wedge b^{\prime}=0,$ and $b\vee b^{\prime}=1$

\item $b<c\Rightarrow c^{\prime}<b^{\prime}$
\end{enumerate}
\end{definition}

\begin{definition}
[from \cite{Beltrametti1981}]A lattice $L$ is called $\sigma$--orthocomplete
when it is orthocomplemented and there exists in $L$ the join of every
countable orthogonal subset of $L$. Furthermore, $L$ is called
\textbf{orthomodular} if, in addition to being $\sigma$--orthocomplete, it
satisfies the condition: $\forall a,b\in B:a\leqslant b\Rightarrow b=a+(b-a)$.
\end{definition}

\begin{definition}
[from \cite{Beltrametti1981}]A lattice $L$ is \textbf{complete} if the meet
and join of any subset of $L$ exist.
\end{definition}

\begin{definition}
[from \cite{Beltrametti1981}]Let $L$ be a lattice. The elements $a,b,c\in L$
form a \textbf{distributive triple} if

\begin{enumerate}
\item $a\wedge(b\vee c)=(a\wedge b)\vee(a\wedge c)$

\item $a\vee(b\wedge c)=(a\vee b)\wedge(a\vee c)$
\end{enumerate}

\noindent with the other four equalities being obtained by cyclical
permutation of $a$, $b$ and $c$.
\end{definition}

\begin{definition}
[from \cite{Beltrametti1981}]A lattice $L$ is \textbf{distributive} if every
triple of elements in $L$\ is distributive.
\end{definition}

\begin{theorem}
[from \cite{Beltrametti1981}]Any distributive lattice is orthomodular.
\end{theorem}

\noindent As mentioned in \cite{Beltrametti1981}, distributivity is a natural
framework for describing classical mechanics, while orthomodularity is the
corresponding framework for quantum mechanics.

The lattice $\mathcal{P(H)}$\ of projectors on a state space $\mathcal{H}$ has
the following characteristics:

\begin{itemize}
\item A projection operator corresponds to a closed subspace of $\mathcal{H}$,
and $\mathcal{P(H)}$ is the set of all closed subspaces of $\mathcal{H}$.

\item $\mathcal{P(H)}$ is partially ordered by set--theoretic inclusion
(denoted $\subseteq$).

\item $\mathcal{P(H)}$ is a complete lattice with a meet and a join.

\item The greatest element of $\mathcal{P(H)}$\ is the whole space
$\mathcal{H}$, and the least element of $\mathcal{P(H)}$\ is the set
consisting only of the zero vector.

\item $\mathcal{P(H)}$ is an orthocomplemented lattice.

\item $\mathcal{P(H)}$ is orthomodular.

\item $\mathcal{P(H)}$ is nondistributive.

\item $\mathcal{P(H)}$ is \emph{atomic} and has the so--called \emph{covering
property. }It is also\emph{ separable.} For an explanation of these terms, see
\cite{Beltrametti1981}.

\item $\mathcal{P(H)}$ is \emph{modular} only if $\mathcal{H}$ is
finite--dimensional. That is why $\mathcal{P(H)}$ is more generally referred
to as \emph{quasi--modular}. Note that there is a subtle difference between
modularity and orthomodularity.
\end{itemize}

The discovery of the structure of $\mathcal{P(H)}$\ is an important result, as
one can reconstruct all the ingredients of the Hilbert--space formalism
(states, transformations and observables) using only the properties of
quantum--mechanical propositions. In \cite{sep-qt-quantlog}, Wilce observes that:

\begin{quotation}
\textsl{From the single premise that the "experimental propositions"
associated with a physical system are encoded by projections in the way
indicated above, one can reconstruct the rest of the formal apparatus of
quantum mechanics. The first step, of course, is Gleason's theorem, which
tells us that probability measures on }$\mathcal{P(H)}$\textsl{ correspond to
density operators. There remains to recover, e.g., the representation of
"observables" by self-adjoint operators, and the dynamics (unitary evolution).
The former can be recovered with the help of the Spectral theorem and the
latter with the aid of a deep theorem of E. Wigner on the projective
representation of groups. [...] The point to bear in mind is that, once the
quantum-logical skeleton }$\mathcal{P(H)}$\textsl{ is in place, the remaining
statistical and dynamical apparatus of quantum mechanics is essentially fixed.
In this sense, then, quantum mechanics -- or, at any rate, its mathematical
framework -- reduces to quantum logic and its attendant probability theory.
\ }
\end{quotation}

\noindent In other words, quantum logic allows us to define the mathematical
framework of quantum mechanics on a more abstract footing.

It is always interesting to find an underlying mathematical relationship
between seemingly unrelated things; in the case of quantum logic, an
investigation into the mathematical structure of quantum mechanics has
revealed the possibility of greatly generalizing propositional logic. We have
only attempted to scratch the surface of this vast and complicated subject
here; however, we have included numerous references for the more advanced reader.%

\paragraph{Bibliographic Notes.}
There are several good textbooks on quantum logic; of these we have selected
and recommend the books by Piron \cite{Piron1976}, Beltrametti and Cassinelli
\cite{Beltrametti1981}, and Mittelstaedt \cite{Mittelstaedt1978}. Piron's book
attempts to unify the mathematical framework of quantum mechanics with that of
classical mechanics, and it uses the hidden--variables interpretation of
quantum theory. In their book, Beltrametti and Cassinelli provide a
comprehensive presentation of the subject, introducing first the
Hilbert--space formalism and then dealing with the relevant algebraic
stuctures in depth; they also describe how the whole formalism may be
reconstructed using only the results of quantum logic. Mittelstaedt's book is
quite readable, and is comparable to Piron's in terms of coverage. Survey
articles of quantum logic include \cite{chiaragiuntini}, by Dalla Chiara and
Giuntini, also \cite{Svozil1998}, by Svozil, and \cite{sep-qt-quantlog}, by
Wilce. The book \cite{coecke-book} is a collection of articles on a particular
variant, known as \emph{operational} quantum logic. The philosopher Hans
Reichenbach designed a simple, three--valued quantum logic which appeals to
one's intuition about measurement \cite{Reichenbach1998}; his exposition of
the philosophical aspects of quantum theory is very readable and informative.

\section{Logics for Quantum Information Systems}

The increased interest in quantum computation and quantum information in
recent years has made quantum logic very relevant today. Ever since research
in quantum logic was initiated by Birkhoff and von Neumann, physicists were
constantly at odds over what its precise form should be. As we saw in the
previous section, the emphasis was mostly put on the \emph{semantics} of
quantum logic --- in particular, on the mathematical structures that underlie
quantum theory. What is needed today is a means of reasoning formally about
systems with quantum--mechanical components and procedures, namely, a
specialized logic with a formal syntax for describing quantum algorithms,
quantum protocols, and their properties. In order to fulfil this need, one
must design a suitable logic using the top--down approach, rather than
starting from low--level algebraic structures and their properties. In the
words of \cite{Mateus2004a}:

\begin{quotation}
\textsl{It is to be expected that the lattice approach to quantum logic will
play a similar role to the one played by modal algebras in modal logic, by
Heyting algebras in intuitionistic logic, by Boolean algebras in classical
logic, etc. But, as in those cases, the algebraic approach is not the right
source of inspiration for discovering the linguistic ingredients of the
envisaged logic. For instance, modal algebras appeared much later than Kripke
structures, well after the modal language was widely accepted.}
\end{quotation}

P. Mateus and A. Sernadas \cite{Mateus2004,Mateus2004a}, and also R. van der
Meyden and M. Patra \cite{meyden-know,Meyden2003,Patra,Patra2005}, are among
those who have taken up this challenge. Their approaches are fundamentally
different to the one of Birkhoff and von Neumann; both pairs of authors have
designed quantum logics which are extensions of probabilistic logic.

Mateus and Sernadas have used the \emph{exogenous} approach to design a logic
for reasoning about quantum systems. This means that they have kept intact the
classical model of propositional logic as the basis for their logic and simply
augmented it to account for the probabilism inherent in quantum mechanics; in
particular, the semantics of their logic is such that the meaning of a quantum
proposition is given by \emph{a superposition of the meanings of classical
propositions. }So, instead of building their logic atop the algebraic
structures of quantum mechanics, Mateus and Sernadas have used models of
propositional logic as their starting point. Their work is particularly
inspired by the semantics of probabilistic logic, as given in
\cite{fagin-halpern,Nilsson1986}. The name of the logic they have proposed is
\textquotedblleft Exogenous Quantum Propositional Logic\textquotedblright%
\ (EQPL). A more powerful version of the logic, which allows one to describe
the dynamics of quantum systems, is \textquotedblleft Dynamic Exogenous
Quantum Propositional Logic\textquotedblright\ (DEQPL).

Van der Meyden and Patra have focused on adapting the probabilistic logic in
\cite{fagin-halpern} to quantum systems, and they have come up with a logic
for knowledge and time in quantum systems \cite{meyden-know}, and a logic for
probability in quantum systems \cite{Meyden2003}. We will only consider the
former of the two logics \cite{meyden-know} here.

\subsection{Exogenous Quantum Propositional Logic (EQPL)}

EQPL \cite{Mateus2004,Mateus2004a} is designed to allow one to write
assertions about quantum systems consisting of a finite number of qubits. The
constructs of the logic allow one to reason about a wide range of systems,
ranging from entangled pairs of qubits to whole quantum cryptographic
protocols. EQPL only allows one to reason about quantum states and
measurements; the extended version of the logic, DEQPL, can be used to write
formulae which include quantum operators.

A quantum system is described in EQPL by a finite set of propositional
constants, $P=\{\mathbf{p}_{k}\;|\;k\in%
\mathbb{N}
\}.$ Each propositional constant $\mathbf{p}_{k}$\ corresponds to a single
qubit in the system under consideration. We define a set $V$ of so--called
\emph{classical valuations} on \emph{P}. A classical valuation is just a
function which attaches a truth value to a propositional constant, so
$V=\left\{  v\;|\;v:P\mapsto\{0,1\}\right\}  $. Classical valuations give
meaning to \textquotedblleft classical formulae\textquotedblright\ which are
used in EQPL. The formal syntax of classical formulae is given in BNF below.%
\begin{equation}
\varphi::=\mathbf{p}_{k}\;|\;(\lnot\varphi)\;|\;(\varphi\Rightarrow\varphi)
\tag{Classical formulae}%
\end{equation}

\noindent In practice, of course, classical formulae in EQPL will also include
other Boolean connectives, such as `and' ($\wedge$), and `or' ($\vee$).
Classical formulae have their usual meaning from propositional logic.

The full language of EQPL includes general formulae, classical formulae, real
terms and complex terms. The syntax of the full language is defined as follows:%

\begin{align}
\gamma &  ::=\varphi\;|\;(t\leqslant t)\;|\;([S]\,\Diamond\,\overrightarrow
{\psi:u})\;|\;(\boxminus\gamma)\;|\;(\gamma\sqsupset\gamma)\tag{Formulae}\\
t  &  ::=r\;|\;(%
\textstyle{\int}%
\varphi)\;|\;(%
\textstyle{\int}%
\varphi_{2}\,|\,\varphi_{1})\;|\;(t+t)\,\;|\;(t\,t)\;|\;\operatorname{Re}%
(u)\;|\;\operatorname{Im}(u)\;|\;\arg(u)\;|\;\left\Vert u\right\Vert \tag{Real
terms}\\
u  &  ::=(t+\mathbf{i}t)\;|\;t\exp(\mathbf{i}t)\;|\;u\;|\;(u+u)\;|\;(u\,u)
\tag{Complex terms}%
\end{align}

\noindent In the above, $r$ denotes a real number, and $\mathbf{i}=\sqrt{-1}$.

The most important ingredients of the logic are (see
\cite{Mateus2004,Mateus2004a} for the formal definitions):

\begin{description}
\item[{$([S]\,\Diamond\,\overrightarrow{\psi:u})$}] Quantum modality; this is
for making assertions about qubits.

\item[$(\boxminus\gamma)$] Quantum \textquotedblleft
negation.\textquotedblright

\item[$(\gamma\sqsupset\gamma)$] Quantum \textquotedblleft
implication.\textquotedblright

\item[$(%
\textstyle{\int}%
\varphi)$] gives the probability of getting an outcome for which $\varphi$
holds, when a measurement is made.

\item[$(%
\textstyle{\int}%
\varphi_{2}\,|\,\varphi_{1})$] gives the probability of getting an outcome for
which $\varphi_{2}$ holds, given that $\varphi_{1}$ holds, when we observe the
quantum system.
\end{description}

A very simple example of the use of the logic is the following EQPL
specification, which describes the state of a two--qubit system in the
entangled state $\frac{1}{\sqrt{2}}(\left\vert 00\right\rangle +\left\vert
11\right\rangle )$.%

\[
\left(  \lbrack\mathbf{p}_{0},\mathbf{p}_{1}]\,\Diamond\;(\mathbf{p}_{0}%
\wedge\mathbf{p}_{1}):\frac{1}{\sqrt{2}},\;((\lnot\mathbf{p}_{0})\wedge
(\lnot\mathbf{p}_{1})):\frac{1}{\sqrt{2}}\right)
\]

\noindent In this specification, $\mathbf{p}_{0}$ and $\mathbf{p}_{1}$ are the
propositional constants corresponding to the two qubits in the system in
question. Obviously, this particular system is uninteresting, but it serves to
illustrate EQPL's syntax. The specification entails the following formulae:

\begin{description}
\item[$(%
\textstyle{\int}%
\mathbf{p}_{0})=\frac{1}{2}$] i.e. \textquotedblleft the probability that,
\emph{the outcome of measuring the first qubit is the truth value 1,} is
$\frac{1}{2}.$\textquotedblright

\item[$(%
\textstyle{\int}%
\mathbf{p}_{1}|\mathbf{p}_{0})=1$] i.e. \textquotedblleft the probability
that, \emph{the outcome of measuring the second qubit is the truth value 1,
given that the outcome of measuring the first qubit was the truth value 1,} is
1.\textquotedblright
\end{description}

Dynamic EQPL (or DEQPL) introduces means of reasoning about state transitions
of a quantum system. It adds to the language of EQPL a number of unitary
quantum operators, as well as notation for projectors. Dynamic EQPL has enough
expressive power to describe quantum protocols, as demonstrated in
\cite{Mateus2004} for the BB84 quantum cryptographic protocol
\cite{nikos-acmCR,nielsenchuang}.

\subsection{A Logic for Knowledge and Time in Quantum Systems}

R. van der Meyden and M. Patra have proposed a modal logic for knowledge and
time in quantum protocols \cite{meyden-know}. They recognise the fact that, in
the literature on quantum computation and information, epistemic locutions of
the form

\begin{quote}
\textquotedblleft\ Alice knows $x$. \textquotedblright
\end{quote}

\noindent are frequently encountered; the logical framework which they propose
is essentially an attempt to make such informal language precise. Their
ultimate objective is to lay the foundations for \textquotedblleft epistemic
analysis\textquotedblright\ of quantum cryptographic protocols, and related
schemes, using logical methods. Here, we will state the syntax of this
\textquotedblleft KT\textquotedblright\ quantum logic and show how it has been
used to specify certain properties of the B92 protocol for quantum key distribution.

The KT quantum logic involves formulas over a set of uninterpreted
propositions, \texttt{Prop}. A formula in the logic may be a proposition, a
conjunction or negation of formulae, or one of the following:

\begin{itemize}
\item the form $\square\phi_{1}$, which retains the usual temporal meaning
(\textquotedblleft always, formula $\phi_{1}$ holds\textquotedblright);

\item the form $\mathit{init}(\phi_{1})$, which is true if $\phi_{1}$ holds in
the initial state of a protocol;

\item the form $K_{i}^{c}(\phi_{1})$, which means \textquotedblleft agent $i $
knows, given her classical bits and observations, that $\phi_{1}$ holds in the
current state;

\item the form $K_{i}^{q}(\phi_{1})$, which means \textquotedblleft agent $i $
knows, given a set of qubits in her possession, that $\phi_{1}$ holds in the
current state.
\end{itemize}

\noindent So, the syntax of formulae, $\Phi$, in the quantum logic is given by
the following grammar:
\[
\Phi::=p\text{ }|\text{ }\phi_{1}\text{ }|\text{ }\phi_{2}\text{ }|\text{
}\phi_{1}\wedge\phi_{2}\text{ }|\text{ }\lnot\phi_{1}\text{ }|\text{ }%
\square\phi_{1}\text{ }|\text{ }\mathit{init}(\phi_{1})\text{ }|\text{ }%
K_{i}^{c}(\phi_{1})\text{ }|\text{ }K_{i}^{q}(\phi_{1})
\]

\noindent where $p\in$\texttt{Prop}. The concept of \textquotedblleft
knowledge\textquotedblright\ has two variations in the logic, since it depends
on what information is used by a particular agent to decide her actions. There
is a concept of \textquotedblleft classical knowledge,\textquotedblright%
\ which is obtained from only classical bits, and a concept of
\textquotedblleft quantum knowledge,\textquotedblright\ which represents
information which can be inferred from a finite set of given quantum states.

In order to define a property using this logic, a model of the protocol under
consideration must be built. The logic assumes that protocols are described as
\emph{qubit message passing environments,} which are defined as follows (we
have modified the original definition slightly):

\begin{definition}
A \textbf{qubit message passing environment} is an abstract model of the
computational setting in a quantum protocol, involving agents and channels for
synchronous communication. It is defined as a tuple
\[
\left\langle n,S,I,Act\right\rangle
\]

\noindent where $n$ is the number of agents involved in the system,
$S=S^{q}\times S^{c}$ is the set of all states that occur in the system, $I$
is the initial state and Act is the set of actions performed by the various agents.
\end{definition}

The global state $S$ is partitioned into a set of classical states, $S^{c}$,
and a set of quantum states, $S^{q}$. Clearly $S^{q}$ is a subset of the
Hilbert space $\mathcal{H}$ of dimension $2^{N}$, the vector space inhabited
by $N$ qubits. The set of classical states consists of elements of the form
$s^{q}=\left\langle var,loc,chan,res\right\rangle $, which include

\begin{itemize}
\item classical bit assignments, $var(i):$ $Var_{i}\longmapsto\{0,1\}$ (here,
$Var_{i}$ is the set of variable names belonging to agent i).

\item qubit location assignments, $loc:[0,N]\mapsto\lbrack0,n]$. The value of
$loc(x)$ is the name of the agent to which $x$ is attached.

\item channel value assignments, $chan:[1..n]^{2}\mapsto$\texttt{Msg}, where
\texttt{Msg} is a set of classical messages. If $chan(i,j)=m$ in a particular
state, this means that message $m\in$\texttt{Msg} has just been transmitted
from agent $i$ to agent $j$.

\item measurement result assignments. If $res(i)=(M^{i},m_{i})$ in a
particular state, it means that the measurement operator $M^{i}$ has been
applied to the quantum states in $S^{q}$, producing as a classical outcome,
the value $m_{i}$.
\end{itemize}

It is instructive to show how the logic can be used to describe certain
properties of the B92 protocol for quantum key distribution
\cite{Bennett92qkd} formally. The B92 protocol is listed in Figure
\ref{b92-meyden-patra}, but the reader should consult one of the many
references on quantum cryptography if he or she is unfamiliar with the technique.

Van der Meyden and Patra generally treat \emph{protocols} as functions
$\mathbf{P}$ pertaining to a particular environment.

\begin{definition}
A \textbf{run} $r:%
\mathbb{N}
\mapsto S$ describes a potential evolution of the system, with $r(m)$
representing the global state of the system at time $m$.
\end{definition}

\begin{definition}
A \textbf{protocol} is a system comprised of specific sets of runs, which are
generated by various agents engaging in a particular pattern of behaviour. For
agent $i$, a protocol is defined as a function $\mathbf{P}:\mathcal{O}_{i}%
^{+}\mapsto Act_{i}$, where $\mathcal{O}_{i}^{+}$ is the set of all
observations the agent has made, and $Act_{i}$ is the set of actions performed
by the agent.
\end{definition}

\begin{figure}[tp]%

\ \ 

\emph{The B92 protocol \cite{Bennett92qkd} allows two users, Alice and Bob, to
establish a common secret key, using a single quantum channel and a classical
communication medium, such as a telephone connection. The idea is to prevent
an eavesdropper (\textquotedblleft Eve\textquotedblright) from obtaining the
value of the key, which is a random binary sequence encoded using qubits;
these qubits are transmitted over the quantum channel, and Bob measures each
in order to recover the encoded bit values. According to quantum theory, only
a compatible measurement is guaranteed to recover the correct bit value. If an
incompatible measurement is made (i.e. a measurement with respect to a
different basis of the qubit's state space), then the correct bit will only be
obtained with probability 0.5. The first part of the protocol, which involves
Alice sending to Bob a sequence of qubits over the quantum channel, is as
follows:}

\begin{enumerate}
\item \textbf{Initial State:} Alice has a single qubit, and a classical bit,
$a$. Bob has two classical bits, $a^{\prime}$ and $b$. The bases for the set
of quantum states $S^{q}$ in the system are $\boxplus=\{\left\vert
0\right\rangle ,\left\vert 1\right\rangle \}$ and $\boxtimes=\{\left\vert
+\right\rangle ,\left\vert -\right\rangle \}$.

\item \textbf{Alice flips her bit, }$a$\textbf{.}

\begin{itemize}
\item If $a=0$, she prepares her qubit in state $\left\vert 0\right\rangle $.

\item If $a=1$, she prepares her qubit in state $\left\vert +\right\rangle $.
\end{itemize}

\item \textbf{Alice transmits her qubit to Bob.}

\item \textbf{Bob flips his bit }$a^{\prime}$\textbf{.}

\begin{itemize}
\item If $a^{\prime}=0$, he measures the qubit with basis $\boxplus.$

\item If $a^{\prime}=1$, he measures the qubit with basis $\boxtimes.$
\end{itemize}

\item \textbf{If the result of the measurement is either }$\left\vert
0\right\rangle $\textbf{\ or }$\left\vert +\right\rangle $\textbf{, Bob sets
}$b=0$\textbf{.} Otherwise, he sets $b=1$.

\item \textbf{Bob sends a classical message to Alice stating the value of }%
$b$\textbf{.}

\item \textbf{The run of the protocol is deemed successful only if }%
$b=1$\textbf{.}
\end{enumerate}

We write $\mathbf{P}$ for the eavesdropping version of B92, in environment
$\mathcal{E}$, if it prescribes the above behaviour for Alice and Bob, and an
eavesdropper, Eve, receives the qubit transmitted as well as Bob's classical
message. For details, consult \cite{meyden-know}. We use the notation
$k_{i}^{x}(a)\equiv K_{i}^{x}(a=0)\vee K_{i}^{x}(a=1)$, where $x\in\{c,q\}$,
to define the properties of $\mathbf{P}$.

\ \ %

\caption
{A simplified model of the B92 protocol, as used by Van der Meyden and Patra to define the protocol's properties
in the quantum logic.}
\label{b92-meyden-patra}
\end{figure}%

The B92 protocol satisfies the following formulae of KT\ quantum logic:

\begin{description}
\item[$\square(b=1\Rightarrow k_{A}^{c}(a)\wedge k_{B}^{c}(a))$] i.e.
\textquotedblleft In successful runs, Alice and Bob come to `classically know'
bit $a.$\textquotedblright

\item[$\square(b=1\Rightarrow\lnot k_{E}^{c}(a))$] i.e. \textquotedblleft Eve
never comes to know bit $a$ based on `classical observations'
alone.\textquotedblright

\item[$\square(b=1\Rightarrow k_{E}^{q}(a))$] i.e. \textquotedblleft If Eve
could perform repeatable measurements on the qubit intercepted, she could come
to learn the value of $a$.\textquotedblright
\end{description}

\section{Concluding Remarks}

Our goal in this brief survey has been primarily to stimulate interest and
provoke thought; we have only attempted to introduce the reader to the
interesting issues at the intersection, so to speak, of logic and quantum
theory. We have introduced the subject of quantum logic and given a brief
account of the literature. We have also given a summary of recent work on
developing logics for quantum information systems.

It is hoped that an understanding of quantum logic will be useful in the quest
to understand and model the structure of Nature's laws, and that computer
scientists will be able, in their own way, to contribute to this adventure.%

\end{document}